# Planar Hexacoordinate Silicon

Chen Chen,[a] Meng-hui Wang,[a] Sudip Pan*[b] and Zhong-hua Cui,*[a]

[a]*Institute of Atomic and Molecular Physics, Key Laboratory of Physics and Technology for Advanced Batteries (Ministry of Education), Jilin University, Changchun, China*

E-mail: zcui@jlu.edu.cn

[b]*Fachbereich Chemie, Philipps-Universität Marburg, Hans-Meerwein-Strasse 4, 35032 Marburg, Germany*

E-mail: pans@chemie.uni-marburg.de

**Abstract**

The occurrence of planar hexacoordination is very rare in cluster chemistry, particularly for main group elements. We report here a class of planar hexacoordinate silicon (phSi) in the global minimum isomer of $SiE_3M_3^+$ (E = N, P, As, Sb; M = Ca, Sr, Ba). Three Si-E multiple bonds between phSi and E centers is a key structural and electronic prerequisite for the observation of their prefect planarity and high stability. Especially, the electrostatic interactions between phSi and three M centers become less repulsive with decrease in electronegativity of E. Eventually, a sizable electrostatic attractive interaction exists in between phSi and M centers in $SiSb_3M_3^+$, leading to a true unprecedented phSi bonding motif which features three Si-Sb multiple bonds and three Si-M ionic bonds.

**Introduction**

Carbon plays the central role to develop the field of hypercoordination in cluster chemistry. Because of remarkable advancement in this field, now the present status could be fairly described as planar teracoordinate carbon (ptC)[1-22] is quite achievable (at least in paper), while planar pentacoordinate carbon (ppC)[23-29] is a tougher target. But planar hexacoordinate carbon (phC) is an extraordinary and still almost unimaginable candidate. The long absence of examples of the latter category in the literature, despite huge advancement in both experimental techniques and computational approaches, supports this argument.

The search for phC was triggered by the report of a perfectly $D_{6h}$-symmetry $CB_6^{2-}$ in 2000,[30] which was later turned out as a just local minimum, located at 30 kcal/mol above the lowest energy isomer.[31] Then, it takes till 2012 to get the first genuine global minimum cluster having six connectivity with carbon in planar form in $D_{3h}$ symmetric $CO_3Li_3^+$ cluster, although electrostatic repulsion between positively charged phC and Li centers and the absence of any significant orbital interaction between them make this hexacoordinate assignment questionable.[31] In a recent study, Tiznado and Merino, and co-workers played with the combination in $CO_3Li_3^+$ cluster and finally succeeded to get a series of phC systems, $CE_3M_3^+$ (E = S, Se, Te; M = Li-Cs) where the natural charge on phC center is negative, and thus an electrostatic attraction exists between phC and positively charged alkali atoms.[32] Note that the understanding of nature of electrostatic interaction based on point charges could be misleading as this is the electronic charge distribution that should be taken into account. This becomes obvious by looking at the electrostatic interaction between two neutral atoms such as in neutral $N_2$. One would expect that the Coulomb interaction is very weak or slightly repulsive in nature, but it is actually highly attractive.[30] The authors in their study performed interacting quantum atoms (IQA) analysis,[31-34] which provides energy components for each individual bonds and it also gives attractive electrostatic interaction between phC and M centers. Therefore, in $CE_3M_3^+$ clusters phC is covalently bonded to three chalcogens and ionically connected to the three alkali metals. Since IUPAC definition[35] of coordination number does not demand orbital interaction, this is fair to call them phC systems.

Now, the question is whether we can also propose viable planar hexacoordinate silicon (phSi). In comparison to carbon, designing of its heavier homologues is more challenging task. This is because although due to the larger size the Steric repulsion between ligands decreases, the bonding strength of π and σ bonds between central atom and peripheral centers also greatly decreases. This is the reason why only a few limited numbers of ppSi and ppGe systems are reported so far.[36] Herein we made efforts to find the correct combination to get phSi system as the most stable isomer. Gratifyingly, we found a series of phSi as global minimum in $SiE_3M_3^+$ (E = N, P, As, Sb; M = Ca, Sr, Ba) clusters through a thorough potential energy surface (PES) exploration. However, the electrostatic interactions between phSi and three M centers are repulsive for E = N and this repulsive nature gradually decreases with reduction in electronegativity of E. Eventually, it becomes attractive in nature in $SiSb_3M_3^+$ cluster.

**Computational methods**

The PES exploration of $SiE_3M_3^+$ (E = N, P, As, Sb; M = Ca, Sr, Ba) was performed using the particle swarm optimization (PSO) algorithm as implemented in the CALYPSO code.[37,38] The random structures generated were initially optimized at the PBE0[39]/def2-SVP level, and then further optimization followed by the harmonic vibrational frequency calculation of the low-lying energy isomers were done at the PBE0/def2-TZVP level. The energies were further refined with the single-point energy calculations for the low-lying energy isomers at the CCSD(T)[40]/def2-TZVP//PBE0/def2-TZVP level. The $T_1$ diagnostic[41] [ref] values are in a reasonable range, indicating that single-reference based method can be used with confidence (see supporting information). Total energies were corrected by the zero-point energies (ZPE) of PBE0. The BO-MD[42] (Born–Oppenheimer molecular dynamics) simulations at 400 K were computed at the PBE0/def2-SVP level. The electronic analysis of global phSi was performed by the natural bond orbital (NBO)[43] and adaptive natural density partitioning (AdNDP)[44] analyses. The iso-chemical shielding surface (ICSS)[45] and the quantum theory of atoms in molecules (QTAIM)[46] analysis were carried out using the

Multiwfn program.[47] All these calculations above were performed using the GAUSSIAN 09 package.[48] The IQA[49] analysis was carried out using the ADF package.[50]

**Structures and stability**

At the first step, we checked whether the $D_{3h}$ symmetric phSi isomer of SiE$_3$M$_3^+$ (E = group 15 element, and M = group 2 element) is a minimum on the PES. The results in Figure S1 show that for E = Be and Mg, phSi isomer possesses a large out-of-plane imaginary frequency mode and thus they are discarded from further consideration. In case of SiBi$_3$M$_3^+$, even for M = Ca the phSi isomer possesses a small imaginary frequency. Although SiBi$_3$Sr$_3^+$ and SiBi$_3$Ba$_3^+$ are true minima, they are 2.2 and 2.5 kcal/mol higher in energy than the lowest energy isomer, respectively (Figure S2). Figure 1 displays some low-lying energy isomers of SiE$_3$M$_3^+$ (E = N, P, As, Sb; M = Ca, Sr, Ba) (see Figures S3-S6 for more isomers). For all the twelve cases, the most stable isomer is a $D_{3h}$ symmetric phSi with $^1$A$_1'$ electronic state. The second lowest energy isomer, which is a ppSi, is located more than 49 kcal/mol above the phSi for E = N. Th energy difference gradually decreases in moving from E = N to E = Sb. In case of SiSb$_3$M$_3^+$ the second lowest energy isomer is 4.6-6.1 kcal/mol higher in energy than phSi. The nearest triplet state isomer is very higher in energy (24-53 kcal/mol) than the global minimum.

BO-MD simulations at 400 K taking SiE$_3$Ca$_3^+$ as case studies were also performed and the related results are displayed in Figure S7. All trajectories show no isomerization or other structural alterations during the simulation time as indicated by small periodic RMSD values (root mean square deviation). This suggests that the global phSi SiE$_3$M$_3^+$ clusters also have reasonable kinetic stability against isomerization.

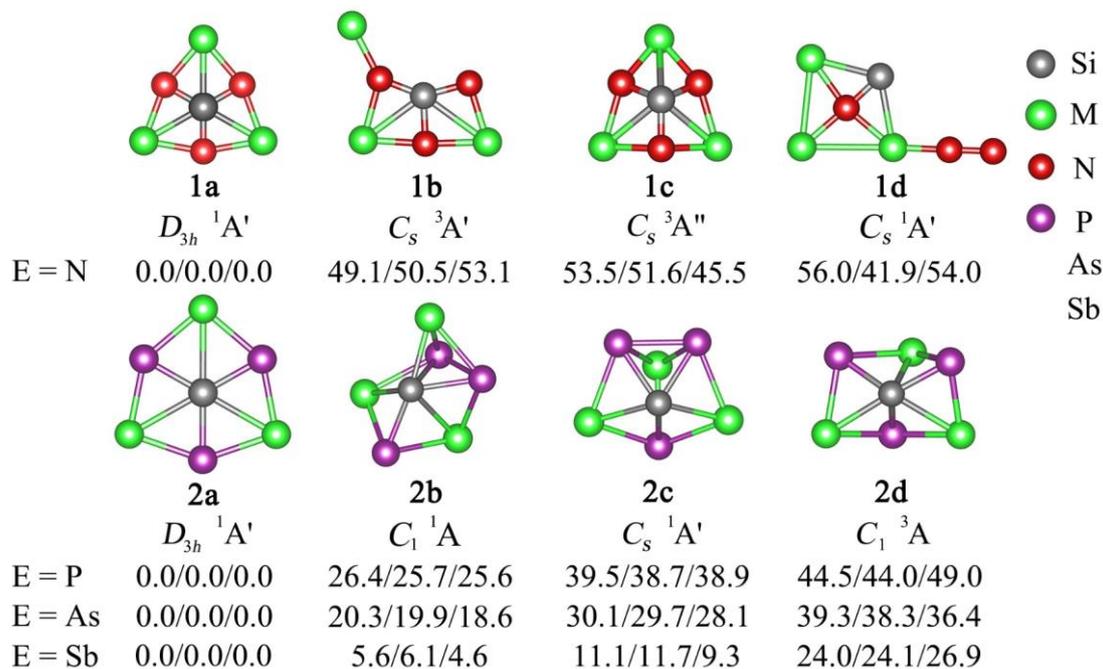

**Figure 1**. The structures and relative energies in kcal/mol of the low-lying energy isomers of $SiE_3M_3^+$ (E=N, P, As, Sb; M=Ca, Sr, Ba) obtained at the single-point CCSD(T)/def2-TZVP//PBE0/def2-TZVP calculations followed by the zero-energy correction of PBE0, the values from left to right refer to Ca, Sr, and Ba in sequence. The group symmetries and spectroscopic states are given.

The bond distances, natural partial charges and Wiberg bond indices (WBIs) for $SiE_3Ca_3^+$ are given in Table 1 (see Tables S1 and S2 for M = Sr, Ba). The Si-E bond distances are shorter than typical Si-E single bond distance computed using the self-consistent covalent radii proposed by Pyykkö,[51] whereas the Si-M bond distance is almost equal to the Si-M single bond. The WBI values for the Si-E bonds lie within 1.1-1.3, while the same for Si-M bonds is almost negligible (0.02-0.04). This indicates partial double bond character in Si-E bonds and very little covalent character in Si-M bonds.

**Table 1.** Bond properties (B, distances in Å and WBIs in parenthesis) and NPA charges (Q, |e|) of $SiE_3Ca_3^+$ (E = N, P, As, Sb) computed at the PBE0/def2-TZVP level.

|        | $B_{Si-E}$ | $B_{Si-Ca}$ | $B_{E-Ca}$ | $Q_{Si}$ | $Q_E$ | $Q_{Ca}$ |
|--------|-----------|-------------|------------|----------|-------|----------|
| E = N  | 1.669 (1.14) | 2.555 (0.02) | 2.246 (0.22) | 1.57 | -1.93 | 1.74 |
| E = P  | 2.180 (1.34) | 2.935 (0.03) | 2.640 (0.27) | 0.25 | -1.42 | 1.67 |
| E = As | 2.301 (1.33) | 3.004 (0.03) | 2.721 (0.29) | 0.07 | -1.34 | 1.65 |
| E = Sb | 2.538 (1.29) | 3.155 (0.04) | 2.896 (0.30) | -0.39 | -1.16 | 1.62 |

The bonding pattern of phSi was uncovered by the AdNDP[44] (adaptive nature density partitioning) analysis. As shown in Figure 2, the AdNDP bonds can be divided in three parts. First, in the first row, the six 1c-2e bonds (one center-two electrons) with occupation number (ON) of 1.71~1.94 are the s/p lone pairs of E atoms (E=N, P, As, Sb). The three Si-E 2c-2e σ bonds (ON=ca. 2.0) is also given in the first row. The 4c-2e π bonds of the second row can be attributed to three π bonds that mainly locate at the SiE$_3$ region, thus combining with Si-E 2c-2e σ bonds reveal that the Si-E bonds possesses a clear multiple bonding characteristic, being consistent with the bonding properties discussed above. These Si-E multiple bonds pattern vividly follows our π-localization strategy, where the SiE$_3$ motifs having a firm π-localized framework provides a driving force for the title phSi planarity and stability.

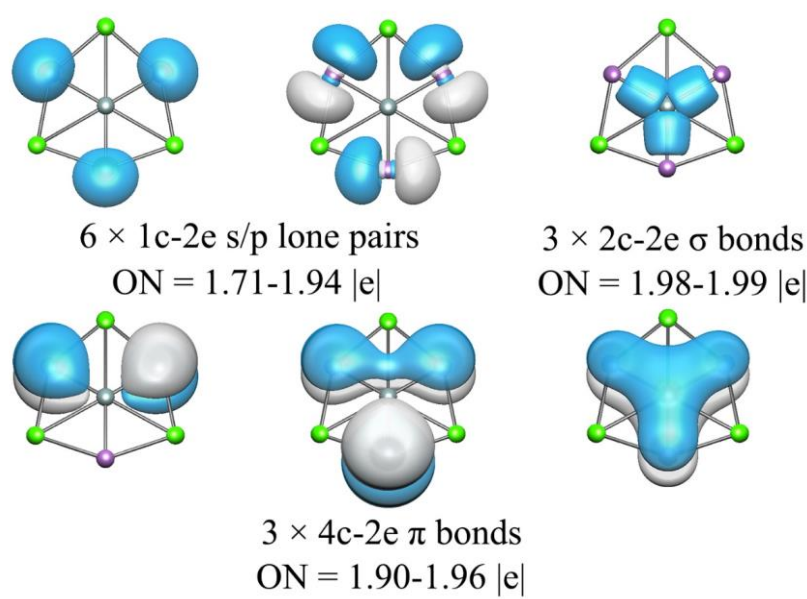

6 × 1c-2e s/p lone pairs
ON = 1.71-1.94 |e|

3 × 2c-2e σ bonds
ON = 1.98-1.99 |e|

3 × 4c-2e π bonds
ON = 1.90-1.96 |e|

**Figure 2.** AdNDP bonds and their occupation numbers (ONs) in |e| of SiE$_3$Ca$_3^+$ (E = N,

P, As, Sb) obtained at the PBE0/def2-TZVP level.

To understand the electronic properties of phSi, the natural bonding orbital (NBO)[43] analysis was performed at the PBE0/def2-TZVP level. As shown in Table 1, the natural population analysis (NPA)[52] charges suggest that the strong charge transfer from the peripheral M ligands to $SiE_3$ motifs occurs in all phSi species, indicated by that the $M^{\delta+}$ with ca. +1.7 |e| but E possesses the large negative charges (-1.2~-1.9 |e|). Such charge distribution reveals that the strong electrostatic attraction accounts for the short E-M contacts, meanwhile, where the bridging $E^{\delta-}$ is responsible for the Si-M short contacts. For the $SiE_3$ motifs, the phSi center has a large positive NPA charge (+1.57 |e|) in $SiN_3Ca_3^+$, but it decreases sharply as it goes from E=N to Sb of $SiE_3Ca_3^+$. Especially, phSi carries a considerable negative charge (-0.39 |e|) in $SiSb_3Ca_3^+$. Thus, the electrostatic interactions between phSi center and three M ligands become less repulsive along with the less electronegative E, and eventually turn into electrostatic attractive occurs in $SiSb_3M_3^+$, making a true planar hexacoordinate silicon bonding. Such electrostatic interaction change behavior is similar to $CE_3Li_3^+$ (E=group 16 elements) where electrostatic interactions between phC and Li gradually become less repulsive along with the less electronegative E, and even turn into electrostatic attraction in global $CSe_3Li_3^+$ and $CTe_3Li_3^+$ clusters.[53] It is worthy of noting that the M ligands is clearly coordinated to phSi or phC although the M-C (phC) or M-Si (phSi) present the electrostatic repulsion according to the coordination definition of IUPAC, where "the coordination number of a specified atom in a chemical species as the number of other atoms directly linked to that specified atom".[54] Nevertheless, the phSi in $SiSb_3Ca_3^+$ and phC in $CSe_3Li_3^+$ and $CTe_3Li_3^+$[53] can be verified to be a true planar hexacoordinate silicon/carbon bonding involving three C(Si)-E multiple bonds and three C(Si)-M ionic bonds.

The interacting quantum atoms (IQA)[49] analysis was considered herein for understanding the inter-atomic interaction energy ($V_{Total}$), which can be evaluated and decomposed into electrostatic (ionic, $V_{Ionic}$) and exchange (covalent, $V_{Coval.}$) contributions. The former is classical electrostatic terms, yet the latter is based on the exchange-correlation part of the electron-electron interaction. The three SiE$_3$M$_3^+$ (M=Ca, Sr, Ba) present an essentially identical behavior of the IQA analysis, thus only SiE$_3$Ca$_3^+$ is given in Table 2 and the other two can be found in Table S3 and S4. As shown in Table 2, the IQA energy decomposition analysis confirms our electronic analysis above. Noteworthy, the Si-Ca inter-atomic interaction displays a repulsion (509.8 kcal/mol) in SiN$_3$Ca$_3^+$ and it sharply decreases in the heavier SiE$_3$Ca$_3^+$ species, and even become attractive in SiSb$_3$Ca$_3^+$, which fully agrees with the results originated from NPA charges distribution in SiSb$_3$Ca$_3^+$, suggesting a true planar hexacoordinate silicon bonding. The Si-N of SiN$_3$Ca$_3^+$ possesses a big interaction energy (V$_{Total}$) of -1232.1 kcal/mol, and it decreases sharply in the heavier SiE$_3$Ca$_3^+$ (E=P, As, Sb) analogues (-450.7 ~ -118.9 kcal/mol), yet the covalent contribution V$_{Coval}$ is gradually enhanced. The similar deceasing behavior from E=N to heavier species also occurs in the E-Ca ligand-ligand interactions (from -476.4 to -185.4 kcal/mol), yet it is essentially ionic. The E-M ligand-ligand stabilizing interactions in all phSi species provide an important driving force between phSi center and M ligands, especially in the SiE$_3$M$_3^+$ (E=N, P, As) cases having a clear electrostatic repulsion. Overall, the $D_{3h}$-symmetry global phSi cluster is stabilized significantly by the three M ligands with additional electrostatic stabilization via the strong charge transfer from M to the SiE$_3$ moiety. Thus, both the Si-E multiple bonds and electrostatic interactions between M ligands to SiE$_3$ make the phSi SiE$_3$M$_3^+$ a highly thermodynamically and kinetically stable species.

**Table 2.** Energy components of interacting quantum atoms (IQA) analysis in kcal/mol for the $D_{3h}$-symmetric phSi SiE$_3$Ca$_3^+$ (E = N, P, As, Sb) systems, where inter-atomic interaction energy (V$_{Total}$) can be evaluated and decomposed into electrostatic (ionic, V$_{Ionic}$) and exchange (covalent, V$_{Coval.}$) contributions.

| SiE$_3$Ca$_3^+$ | SiN$_3$Ca$_3^+$ | SiP$_3$Ca$_3^+$ | SiAs$_3$Ca$_3^+$ | SiSb$_3$Ca$_3^+$ |
|---|---|---|---|---|
| V$_{Total}$ (Si-E) | -1232.1 | -450.7 | -247.6 | -118.9 |
| V$_{Ionic}$ (Si-E) | -1117.6 | -303.5 | -95.1 | 25.2 |
| V$_{Coval.}$ (Si-E) | -114.5 | -147.1 | -152.5 | -144.1 |
| V$_{Total}$ (Si-Ca) | 509.8 | 194.7 | 80.7 | **-40.2** |
| V$_{Ionic}$ (Si-Ca) | 515.2 | 202.9 | 90.6 | **-28.6** |

| $V_{Coval.}$ (Si-Ca) | -5.5 | -8.2 | -10.0 | **-11.6** |
| $V_{Total}$ (E-Ca) | -476.4 | -303.9 | -252.7 | -185.4 |
| $V_{Ionic}$ (E-Ca) | -413.1 | -249.8 | -201.6 | -136.1 |
| $V_{Coval.}$ (E-Ca) | -63.3 | -54.1 | -51.1 | -49.4 |

The bonding pattern and electronic properties of the title phSi based on π-localization strategy is clearly different from the planar species using the π-delocalization strategy,[4] where the latter possesses the balance but weak planar center-ligand bonds, yet the strong covalent bonds occur in the former cases. Beside bonding pattern, the planar hypercoordinate center possesses a different picture, the valence population (VP) of the title phSi atom in $SiE_3Ca_3^+$ is $3s^{0.59\sim1.35}3p_x^{0.53\sim1.00}3p_y^{0.53\sim1.00}3p_z^{0.63\sim0.95}$, where 3s and 3p occupancies gradually increase as it goes from E=N to E=Sb (see Table S5). However, these occupancies are significantly lower than the prototype π-delocalization ptC, for example, $CAl_4^{2-}$ ($2s^{1.62}2p_x^{1.77}2p_y^{1.77}2p_z^{1.61}$).[4,21] Such low valence occupancies are due to the strong charge transfer from phSi center to ligand E, yet the strong Si-E multiple bonds make phSi center a considerable total WBI value (3.5~4.1).

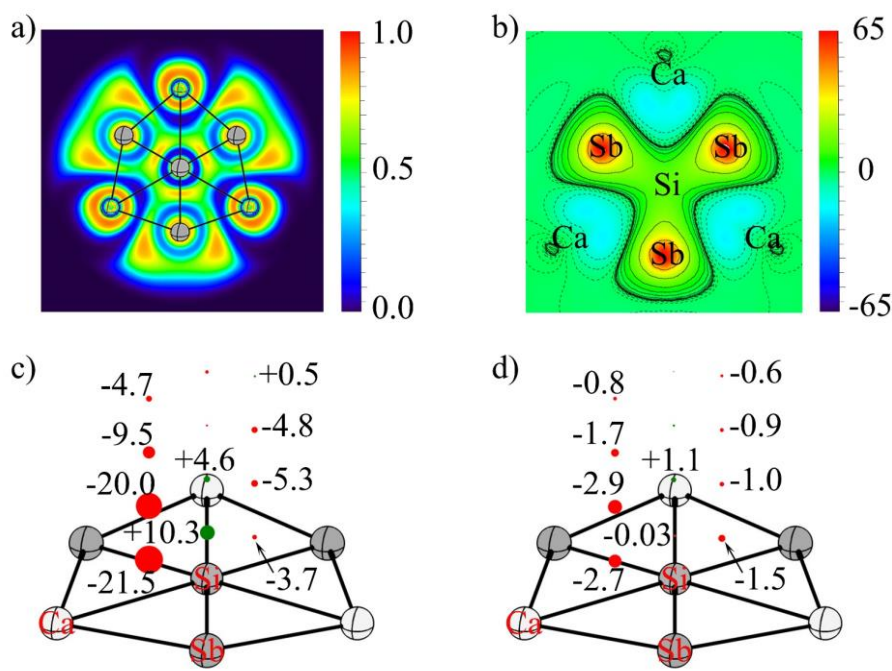

**Figure 3.** a) ELF, b) $ICSS_{zz}$, c) $NICS_{zz}$, and d) $NICS_{zz}$ (π orbitals) for $SiSb_3Ca_3^+$. In d)

the diatropic and paratropic tensors are shown in red and green, respectively. NICS values are in ppm All the cases see Figure S8.

In contrast to the pronounced aromaticity because of π delocalization,[4] the essential feature in π-localized planar species is that π electrons mainly localize at strong covalent bonds rather than delocalize the whole plane, eventually leading to a relatively weak aromaticity. In details, the aromaticity character of the title phSi clusters is uncovered by the iso-chemical shielding surface (ICSS)[45] that is the isosurface of the nucleus-independent chemical shift (NICS) and presents an exceedingly intuitive picture on aromaticity. The $ICSS_{zz}(1)$ (a z-direction component of $ICSS_{zz}$ at 1.0 Å above) is displayed at Figure 3b and Figure S8. The phSi $SiE_3Ca_3^+$ clearly displays a weak aromaticity for the whole plane, featuring a sharp change from Si-E bonds to the neighboring region, indicating -20.0 ppm at Si-Sb region to +4.6 ppm (Si-Ca) and -5.3 ppm (Sb-Si-C triangle) as given Figure 3c. The canonical molecular orbital (CMO) dissection into π orbitals of $NICS_{zz}$ grid is given in Figure 3d, where $NICS_{zz}$ (π electrons) the clearly smaller as compared to the total $NICS_{zz}$. Additionally, the strong electron localization within Si-E multiple bonds is also found in the electron localization function (ELF) picture in Figure 3a. Overall, the electronic analysis of phSi strongly revealed that strong Si-E multiple bonds rather than π electron delocalization give rise to the extraordinary stability of the title phSi cluster, effectively verifying the assessment of the π-localizated strategy proposed herein.

**Conclusions**

In summary, the π-localized design approach is further extended here and results in the first global isolated planar hexacoordinate silicon. The twelve phSi clusters have been achieved in $SiE_3M_3^+$, in which the Si-E multiple bonds and strong electrostatic interactions between ligand M and $SiE_3$ motifs are effectively formed, giving rise to the high stability and planarity of the title phSi species. The electrostatic interactions between phSi center and three M ligands become less repulsive along with the less electronegative E, and eventually turn into the firm electrostatic attractive in $SiSb_3M_3^+$, making an unprecedented planar hexacoordinate silicon bonding associated with three

Si-Sb multiple bonds and three Si-M ionic bonds. The π-localization strategy not only breaks the limit of π-acceptor/σ-donor ligands, but enriches the number of valence electron (usually 18 electrons) for designing planar species, as well as gives rise to a new insight on the chemical bonding and electronic properties of planar species.

## Acknowledgments

This work was funded by the National Natural Science Foundation of China (No. 11922405, 11874178, 91961204). The partial calculations in this work supported by High Performance Computing Center of Jilin University, China.

## Conflict of interest

The authors declare no conflict of interest.